\documentclass[onecolumn,eqsecnum,pra]{revtex4}

\usepackage{amssymb}

\usepackage{amsmath}

\begin{document}

\title{Quantum thermodynamic functions for an oscillator
coupled to a heat bath}

\author{G. W. Ford}

\affiliation{Department of Physics, University of Michigan, Ann Arbor,
MI\\ 48109-1040}

\author{R. F. O'Connell \footnote{Electronic
address:oconnell@phys.lsu.edu}}

\affiliation{Department of Physics and Astronomy, Louisiana State University,
Baton Rouge, LA 70803-4001}

\date{\today }

\begin{abstract}
Small systems (of interest in the areas of nanophysics, quantum
information, etc.) are particularly vulnerable to environmental effects. 
Thus, we determine various thermodynamic functions for an oscillator in
an arbitrary heat bath at arbitrary temperatures.  Explicit results are
presented for the most commonly discussed heat bath models:  Ohmic, single
relaxation time and blackbody radiation.
\\
\\
\\
PAC Numbers: 05.30.-d, 05.40.-a, 05.70.-a
\\
\\
\\
\\
\\
\end{abstract}

\maketitle

\section{Introduction}

\label{sec:one}

Heat bath models are of topical interest especially in areas such as
quantum information and nanophysics.  Thus, it is important to develop
realistic calculations that can be used to make contact with
experiments.  Here, we wish to examine the effects of a heat bath on
various thermodynamic functions such as entropy, partition function,
average energy, specific heat and heat capacity.  Our starting point is
based on an exact result which we have previously derived for the free
energy of an oscillator in an arbitrary heat bath, in terms of a single
integral involving the generalized susceptibility \cite{ford1} arising
from the associated quantum Langevin equation \cite{ford2}.  This result
was used in a series of papers to obtain free energy shifts of atomic
levels in a blackbody radiation field; the effect of a heat bath on the
magnetic moment of an electron gas
\cite{li}, based on a generalization of our previous work to include a
magnetic field; a proof that the third (Nernst's) law of thermodynamics
is valid in the presence of a heat bath
\cite{ford4} and a demonstration that a supposed violation of the second
law is only apparent \cite{ford5}.  Thus, because of its wide
applicability, we are motivated to systematically develop explicit
results for the most commonly discussed heat bath models.  Hence, in Sec.
II, we review our starting-point Hamiltonian describing an oscillator in an
arbitrary heat bath at temperature $T$ which enabled us to obtain the
equation of motion of the oscillator in terms of a quantum Langevin
equation which, in turn led us to an exact expression for the free energy
of an oscillator in an arbitrary heat bath.  Next, we use this general
result to consider in detail the most commonly discussed heat bath models,
obtaining results for the free energy $F(T)$ which incorporates the Ohmic,
single relaxation time and blackbody radiation models in a form which is
very similar for all cases, involving the Stieltjes J-function, whose
properties we present in Appendix A.  This enables us to obtain a simple
expression, in terms of the J-function for the free energy
$F(T)$ which incorporates the Ohmic, single relaxation time and blackbody
radiation models.  This expression for $F(T)$ is then used to obtain
explicit results, both for low temperature and high temperature, for
various thermodynamic functions such as the specific heat, the energy and
the heat capacity; these results are presented in III (for the Ohmic model)
and IV (for the single relaxation and blackbody radiation models). 
Results for the $T=0$ case are given in Sec. V.  We conclude with a brief
summary and discussion in Sec. VI.

\section{Free energy}

\label{sec:two}

The most general coupling of a quantum particle coupled to a linear
passive heat bath is equivalent to an independent-oscillator model
\cite{ford1,ford2}, which is described by the Hamiltonian
\begin{equation}
H=\frac{p^{2}}{2m}+V(x)+\sum_{j}\left(\frac{p^{2}_{j}}{2m_{j}}+\frac{1}{2}m_{j}\omega^{2}_{j}
\left(q_{j}-x\right)^{2}\right).
\label{eq1}
\end{equation}
Here $x$ and $p$ are the particle coordinate and momentum operators and
$V(x)$ is the potential energy of an external force.  The $jth$
independent oscillator has coordinate $q_{j}$ and momentum
$p_{j}$ and the generality of the model arises from the infinity of
oscillators with an arbitrary choice of the mass $m_{j}$ and frequency
$\omega_{j}$ for each.

Use of the Heisenberg equations of motion leads to the 
quantum Langevin equation
\begin{equation}
m\ddot{x}+\int^{t}_{-\infty}{\textnormal{d}}t^{\prime}\mu
(t-t^{\prime})\dot{x}(t^{\prime})
+V^{\prime}(x)=F(t), \label{eq2}
\end{equation}
where $\mu(t)$ is the so-called memory function.  $F(t)$ is the random
(fluctuation or noise) operator force with mean $\langle F(t)\rangle =0$. 
The quantities $\mu (t)$ and
$F(t)$ describe the properties of the heat bath and are independent of the
external force.

In the particular case of an oscillator potential
\begin{equation}
V(x)=\frac{1}{2}Kx^{2}=\frac{1}{2}m\omega^{2}_{0}x^{2}. \label{eq3}
\end{equation}
Substituting (\ref{eq3}) into (\ref{eq2}) enables us to obtain the
explicit solution

\begin{equation}
x(t)=\int^{t}_{-\infty}~dt^{\prime}G(t-t^{\prime})F(t^{\prime}),
\end{equation} where $G$ is the Green function.  The Green function
vanishes for negative times and its Fourier transform,

\begin{equation}
\alpha(\omega)=\int^{\infty}_{0}~dte^{i\omega{t}}G(t)
\end{equation} is the familiar response function (generalized
susceptibility).  This is given by

\begin{equation}
\alpha (z)=\frac{1}{-mz^{2}-iz\tilde{\mu}(z)+K}, \label{eq6}
\end{equation}
where $\tilde{\mu}(z)$ is the Fourier transform of the memory function:
\begin{equation}
\tilde{\mu}(z)=\int^{\infty}_{0}{\textnormal{d}}t\mu (t)e^{izt}.
\label{eq7}
\end{equation}  Note that $\tilde{\mu}(z)$ and, hence, also $\alpha(z)$
are analytic in the upper half plane.\\

The system of an oscillator coupled to a heat bath in thermal equilibrium
at temperature $T$ has a well-defined free energy.  The free energy
ascribed to the oscillator, $F(T)$, is given by the free energy of the
system minus the free energy of the heat bath in the absence of the
oscillator.  This calculation was carried out by two different methods
\cite{ford2,ford6} leading to the "remarkable formula"
\begin{equation}
F(T)=\frac{1}{\pi}\int^{\infty}_{0}d\omega f(\omega
,T){\textnormal{Im}}\left\{\frac{d~\log\alpha (\omega
+i0^{+})}{d\omega}\right\},
\label{eq9}
\end{equation}
where $f(\omega ,T)$ is the free energy of a single oscillator of
frequency $\omega$, given by
\begin{equation}
f(\omega ,T)=kT\log [1-\exp\left(-\hbar\omega /kT\right)]. \label{eq10}
\end{equation}
Here the zero-point contribution $(\hbar\omega /2)$ has been omitted, but
in a brief Sec. V we remark upon this contribution.  We have referred to
(2.8) as a "remarkable formula"
\cite{ford2,ford6}, in the sense that it displays a non-trivial dependence
on the temperature
$T$, in contrast with the corresponding classical formula.  We have now all
the basic tools at our disposal and we proceed to consider three cases of
interest:

\begin{eqnarray}
\tilde{\mu}(z) &=&\zeta ,\qquad \text{Ohmic,}  \notag \\
\tilde{\mu}(z) &=&\frac{\zeta }{1-iz\tau },\qquad \text{Single relaxation
time,}  \notag \\
\tilde{\mu}(z) &=&\frac{2e^{2}z\Omega ^{2}}{3c^{3}(z+i\Omega )},\qquad
\text{Quantum electrodynamics (QED)}.
\end{eqnarray} Here $\zeta $ is the Ohmic friction constant, while $\tau $ is the
relaxation time. \ It is generally assumed that the relaxation time is small
in the sense that $\tau \ll \zeta /m$. In the QED case, $\Omega $ is a high
frequency cutoff characterizing the electron form factor ($\Omega
\rightarrow \infty $ corresponds to a point electron).  The susceptibility
for all three cases may be combined in a single expression

\begin{equation}
\alpha (z)=\frac{z+i\Omega }{-m(z+i\Omega ^{\prime })(z^{2}+i\gamma
z-\omega _{0}^{2})}.
\end{equation} For the single relaxation time model

\begin{equation}
\tau =\frac{1}{\Omega }=\frac{1}{\Omega ^{\prime }+\gamma },\quad \frac{
\zeta }{m}=\gamma \frac{\Omega ^{\prime 2}+\gamma \Omega ^{\prime }+\omega
_{0}^{2}}{(\Omega ^{\prime }+\gamma )^{2}},\quad \frac{K}{m}=\omega
_{0}^{2}\frac{\Omega ^{\prime }}{\Omega ^{\prime }+\gamma }.
\end{equation} The Ohmic model corresponds to the limit of $\Omega ^{\prime }\rightarrow
\infty $, in which case $\tau \rightarrow {0},\ \zeta /m\rightarrow \gamma $
and $K/m\rightarrow \omega _{0}^{2}$. For the QED model

\begin{equation}
\frac{1}{\Omega }=\frac{1}{\Omega ^{\prime }}+\frac{\gamma }{\omega
_{0}^{2}},\qquad \frac{K}{M}=\omega _{0}^{2}\frac{\Omega ^{\prime
}}{\Omega ^{\prime}+\gamma },\qquad \frac{M}{m}=\frac{(\omega
_{0}^{2}+\gamma \Omega ^{\prime})(\Omega ^{\prime }+\gamma )}{\omega
_{0}^{2}\Omega ^{\prime }},
\end{equation} where $m$ is the bare mass and 

\begin{equation}
M=m+\frac{2e^{2}\Omega }{3c^{3}}
\end{equation}

is the renormalized (observed) mass. In this QED case, the limit $\Omega
^{\prime }\rightarrow \infty $ corresponds to the largest value of the
cutoff $\Omega $ consistent with a positive bare mass, that is, in this
limit $m=0$, $K=M\omega _{0}^{2}$ and $\Omega =1/\tau _{\text{e}}$, where

\begin{equation}
\tau _{e}=\frac{2e^{2}}{3Mc^{3}}=6\times {10}^{-24}s.
\end{equation}

With the general form (2.11) the free energy (\ref{eq9}) can
be written

\begin{equation}
F(T)=\frac{kT}{\pi }\int_{0}^{\infty }d\omega \log (1-e^{-\hbar \omega
/kT})\left( -\frac{\Omega }{\omega ^{2}+\Omega ^{2}}+\frac{\Omega^{\prime
}}{\omega ^{2}+\Omega ^{\prime 2}}+\frac{\omega ^{2}+\omega
_{0}^{2}}{(\omega^{2}-\omega _{0}^{2})^{2}+\gamma ^{2}\omega ^{2}}\right).
\end{equation} We use partial fractions in the third term by introducing

\begin{equation}
z_{1}=\frac{\gamma }{2}+i\omega _{1},\quad z_{1}^{\ast }=\frac{\gamma }{2}
-i\omega _{1},\quad \omega _{1}=\sqrt{\omega _{0}^{2}-\frac{\gamma
^{2}}{4}},
\end{equation} and we note that, for the overdamped case
$[(\gamma/2)>\omega_{0}]$, $\omega_{1}$ is imaginary, in which case
$z_{1}=\frac{\gamma}{2}-|\omega_{1}|$ and
$z_{1}^{\ast}=\frac{\gamma}{2}+|\omega_{1}|$.  Hence

\begin{gather}
F(T)=\frac{kT}{\pi }\int_{0}^{\infty }d\omega \log (1-e^{-\hbar \omega
/kT})\left( -\frac{\Omega }{\omega ^{2}+\Omega ^{2}}+\frac{\Omega ^{\prime
}}{\omega ^{2}+\Omega ^{\prime 2}}\right.  \notag \\
\left. +\frac{z_{1}}{\omega ^{2}+z_{1}^{2}}+\frac{z_{1}^{\ast }}{\omega
^{2}+z_{1}^{\ast 2}}\right) \notag \\
=kT\left\{ J(\frac{\hbar \Omega }{2\pi kT})-J(\frac{\hbar \Omega
^{\prime }}{2\pi kT})-J(\frac{\hbar z_{1}}{2\pi kT})-J(\frac{\hbar
z_{1}^{\ast }}{2\pi kT})\right\},  \label{Free_energy_SRT_QED}
\end{gather}

where $J(z)$ is the Stieltjes J-function

\begin{equation}
J(z)=-\frac{1}{\pi }\int_{0}^{\infty }dt\log (1-e^{-2\pi
t})\frac{z}{t^{2}+z^{2}},\qquad \text{Im}z>0.
\end{equation} In the next two sections, we consider the three specific
models separately and in detail. For this purpose, we make extensive use
of the J-function, whose properties are discussed in detail in Appendix
A.

\section{Ohmic model}

\label{sec:three}

Here

\begin{equation}
F(T)=-kT\left[ J(\frac{\hbar z_{1}}{2\pi kT})+J(\frac{\hbar z_{1}^{\ast
}}{2\pi kT})\right],
\end{equation} where in the expression (2.17) for $z_{1}$ and $z_{1}^{\ast
}$ we put $\omega _{0}=\sqrt{K/m}$ and $\gamma =\zeta /m$.

\subsection{Low temperature expansion ($kT<<\hbar\omega_{0}$)}

In the low temperature case we use the asymptotic expansion (A5) for $
J$. With this we obtain for the free energy 

\begin{eqnarray}
F(T) &=&-kT\sum_{n=0}^{\infty }c_{n}  \notag \\
&=&-\left[ \frac{\pi (kT)^{2}\gamma }{6\hbar \omega _{0}^{2}}+\frac{\pi
^{3}(kT)^{4}\gamma \left( 3\omega _{0}^{2}-\gamma ^{2}\right) }{45\hbar
^{3}\omega _{0}^{6}}+\frac{8\pi ^{5}(kT)^{6}\gamma \left( 5\omega
_{0}^{4}-5\gamma ^{2}\omega _{0}^{2}+\gamma ^{4}\right) }{315\hbar
^{5}\omega _{0}^{10}}+\cdots \right].
\end{eqnarray}  The entropy is

\begin{eqnarray}
S(T) &=&-\frac{\partial F(T)}{\partial T}  \notag \\
&=&k\left[ \frac{\pi kT\gamma }{3\hbar \omega _{0}^{2}}+\frac{4\pi
^{3}(kT)^{3}\gamma \left( 3\omega _{0}^{2}-\gamma ^{2}\right) }{45\hbar
^{3}\omega _{0}^{6}}+\frac{16\pi ^{5}(kT)^{5}\gamma \left( 5\omega
_{0}^{4}-5\gamma ^{2}\omega _{0}^{2}+\gamma ^{4}\right) }{105\hbar
^{5}\omega _{0}^{10}}\cdots \right]
\end{eqnarray}  

\begin{equation*}
+\frac{16\pi ^{5}(kT)^{6}\gamma \left( 5\omega _{0}^{4}-5\gamma ^{2}\omega
_{0}^{2}+\gamma ^{4}\right) }{105\hbar ^{5}\omega _{0}^{10}}-\frac{8\pi
^{5}(kT)^{6}\gamma \left( 5\omega _{0}^{4}-5\gamma ^{2}\omega
_{0}^{2}+\gamma ^{4}\right) }{315\hbar ^{5}\omega _{0}^{10}}.
\end{equation*} The energy is

\begin{eqnarray}
U(T) &=&F+TS  \notag \\
&=&\frac{\pi (kT)^{2}\gamma }{6\hbar \omega _{0}^{2}}+\frac{\pi
^{3}(kT)^{4}\gamma \left( 3\omega _{0}^{2}-\gamma ^{2}\right) }{15\hbar
^{3}\omega _{0}^{6}}+\frac{8\pi ^{5}(kT)^{6}\gamma \left( 5\omega
_{0}^{4}-5\gamma ^{2}\omega _{0}^{2}+\gamma ^{4}\right) }{63\hbar
^{5}\omega _{0}^{10}}+\cdots.
\end{eqnarray}  The specific heat is

\begin{eqnarray}
C(T) &=&T\frac{\partial S}{\partial T}  \notag \\
&=&k\left[ \frac{\pi kT\gamma }{3\hbar \omega _{0}^{2}}+\frac{4\pi
^{3}(kT)^{3}\gamma \left( 3\omega _{0}^{2}-\gamma ^{2}\right) }{15\hbar
^{3}\omega _{0}^{6}}+\frac{16\pi ^{5}(kT)^{5}\gamma \left( 5\omega
_{0}^{4}-5\gamma ^{2}\omega _{0}^{2}+\gamma ^{4}\right) }{21\hbar
^{5}\omega _{0}^{10}}+\cdots \right].
\end{eqnarray}  As a check, we note that the leading term in (3.2) agrees
with the result obtained by us in \cite{ford4} while the leading term in
(3.3) agrees with our earlier results \cite{ford4} as well as a recent
result of Hanggi and Ingold \cite{hang}.  In addition, the first two terms
in (3.5) agree with the results obtained in \cite{hang}.

\subsection{High temperature expansion ($kT>>\hbar\omega_{0}$)}

In the high temperature case we use the small argument expansion (A4)
for $J$, with the result

\begin{eqnarray}
F(T) &=&-kT\log \frac{kT}{\hbar \omega _{0}}-\frac{\hbar \gamma }{2\pi
}\log 
\frac{2\pi kT}{\hbar \omega _{0}}-\frac{\hbar \omega _{1}}{\pi }\arccos 
\frac{\gamma }{2\omega _{0}}-\frac{\hbar \gamma }{2\pi }(1-\gamma _{E}) 
\notag \\
&&-2kT\sum_{n=2}^{\infty }(-)^{n}\frac{\zeta (n)}{n}\left( \frac{\hbar
\omega _{0}}{2\pi kT}\right) ^{n}\cos \left( n\arccos \frac{\gamma
}{2\omega _{0}}\right) .
\end{eqnarray}

As a check we consider the uncoupled oscillator. Forming the limit $\gamma
\rightarrow 0$, we find

\begin{eqnarray}
F(T) &\rightarrow &-kT\log \frac{kT}{\hbar \omega _{0}}-\frac{\hbar \omega
_{0}}{2}+kT\sum_{n=1}^{\infty }\frac{\zeta (2n)}{n}\left( \frac{\hbar
\omega _{0}}{2\pi kT}\right) ^{2n}  \notag \\
&=&kT\log (1-e^{-\hbar \omega _{0}/kT}),
\end{eqnarray} which is the familiar result (2.9) for the uncoupled
oscillator. Here we have used the formula \cite{bate}

\begin{equation}
\log (1-e^{-z})=\log z-\frac{1}{2}+\sum_{n=1}^{\infty
}(-)^{n+1}\frac{\zeta (2n)}{n}\left( \frac{z}{2\pi }\right) ^{n}.
\end{equation}

Returning to the expansion (3.6), we obtain explicit expressions for the
first few terms,

\begin{eqnarray}
F(T) &=&-kT\log \frac{kT}{\hbar \omega _{0}}-\frac{\hbar \gamma }{2\pi
}\log 
\frac{2\pi kT}{\hbar \omega _{0}}-\frac{\hbar \omega _{1}}{\pi }\arccos 
\frac{\gamma }{2\omega _{0}}-\frac{\hbar \gamma }{2\pi }(1-\gamma _{E}) 
\notag \\
&&+\frac{\hbar ^{2}(2\omega _{0}^{2}-\gamma ^{2}).}{48kT}-\frac{\zeta
(3)\hbar ^{3}\gamma (3\omega _{0}^{2}-\gamma ^{2})}{24\pi ^{3}(kT)^{2}}
+\cdots .
\end{eqnarray}

With this, the entropy, energy, and specific heat are given, respectively,
by

\begin{eqnarray}
S(T) &=&-\frac{\partial F(T)}{\partial T}  \notag \\
&=&k(\log \frac{kT}{\hbar \omega _{0}}+1)+\frac{\hbar \gamma }{2\pi T} 
\notag \\
&&-2k\sum_{n=2}^{\infty }(-)^{n}\frac{(n-1)\zeta (n)}{n}\left( \frac{\hbar
\omega _{0}}{2\pi kT}\right) ^{n}\cos \left( n\arccos \frac{\gamma
}{2\omega _{0}}\right) .  \notag \\
&=&k(\log \frac{kT}{\hbar \omega _{0}}+1)+\frac{\hbar \gamma }{2\pi T} 
\notag \\
&&+k\frac{\hbar ^{2}(2\omega _{0}^{2}-\gamma
^{2})}{48(kT)^{2}}-k\frac{\zeta (3)}{12\pi ^{3}}\frac{\hbar ^{3}\gamma
(3\omega _{0}^{2}-\gamma ^{2})}{(kT)^{3}},
\end{eqnarray} 

\begin{eqnarray}
U(T) &=&F+TS  \notag \\
&=&kT-\frac{\hbar \gamma }{2\pi }\left( \log \frac{2\pi kT}{\hbar \omega
_{0}}-\gamma _{E}\right) -\frac{\hbar \omega _{1}}{\pi }\arccos
\frac{\gamma }{2\omega _{0}}  \notag \\
&&-2kT\sum_{n=2}^{\infty }(-)^{n}\zeta (n)\left( \frac{\hbar \omega
_{0}}{2\pi kT}\right) ^{n}\cos \left( n\arccos \frac{\gamma }{2\omega
_{0}}\right)\notag \\
&=&kT-\frac{\hbar \gamma }{2\pi }\left( \log \frac{2\pi kT}{\hbar \omega
_{0}}-\gamma _{E}\right) -\frac{\hbar \omega _{1}}{\pi }\arccos
\frac{\gamma }{2\omega _{0}}  \notag \\
&&+\frac{\hbar ^{2}(2\omega _{0}^{2}-\gamma ^{2})}{24kT}-\frac{\zeta
(3)}{8\pi ^{3}}\frac{\hbar ^{3}\gamma (3\omega _{0}^{2}-\gamma
^{2})}{(kT)^{2}},
\end{eqnarray} and

\begin{eqnarray}
C(T) &=&T\frac{\partial S}{\partial T}  \notag \\
&=&k-\frac{\hbar \gamma }{2\pi T}+2k\sum_{n=2}^{\infty }(-)^{n}(n-1)\zeta
(n)\left( \frac{\hbar \omega _{0}}{2\pi kT}\right) ^{n}\cos \left(
n\arccos\frac{\gamma }{2\omega _{0}}\right)  \notag \\
&=&k-\frac{\hbar \gamma }{2\pi T}-k\frac{\hbar ^{2}(2\omega
_{0}^{2}-\gamma ^{2})}{24(kT)^{2}}+k\frac{\zeta (3)}{4\pi^{3}}\frac{\hbar
^{3}\gamma (3\omega _{0}^{2}-\gamma ^{2})}{(kT)^{3}}.
\end{eqnarray}  Note that all these results apply to the overdamped case
with the prescription

\begin{equation*}
\omega _{1}\arccos \frac{\gamma }{2\omega _{0}}\rightarrow \left\vert \omega
_{1}\right\vert \log (\frac{\gamma }{2\omega _{0}}-\frac{\left\vert \omega
_{1}\right\vert }{\omega _{0}}).
\end{equation*}Also, we again have a check in that the first three terms
in the specific heat agree with the results obtained in \cite{hang} for the
Ohmic model.

\section{Single relaxation time and nonrelativistic QED models}

The free energy is now of the general form (\ref{Free_energy_SRT_QED}),
which can be written

\begin{equation}
F(T)=F_{\text{Ohmic}}(T)+kT\left[ J(\frac{\hbar \Omega }{2\pi
kT})-J(\frac{\hbar \Omega ^{\prime }}{2\pi kT})\right] .
\end{equation}  We argue that $\Omega$ and $\Omega^{\prime}$ will always
be large compared with $kT$, so it is appropriate to use the low
temperature expansion and then only the first term. The result is

\begin{equation}
F(T)=F_{\text{Ohmic}}(T)+\frac{\pi (kT)^{2}}{6\hbar }\left( \frac{1}{\Omega
} -\frac{1}{\Omega ^{\prime }}\right) .
\end{equation} For the single relaxation time case, it is clear from (2.12)
that the second term in (4.2) is very small so that the results in this
case are essentially the same as for the Ohmic case.  However, for the QED
case

\begin{equation}
\frac{1}{\Omega }=\frac{1}{\Omega ^{\prime }}+\frac{\gamma }{\omega
_{0}^{2}},
\end{equation} so

\begin{equation}
F(T)=F_{\text{Ohmic}}(T)+\frac{\pi (kT)^{2}\gamma }{6\hbar \omega
_{0}^{2}}.
\end{equation}

\subsection{Low temperature expansion ($kT<<\hbar\omega_{0}$)}

The second term in (4.4) is exactly the negative of the leading term
in the low temperature expansion $(kT<<\hbar\omega_{0})$ for the Ohmic
case, given in (3.2). In other words, the $T^{2}$ term vanishes and the
leading term is the
$T^{4}$ term.  The result is that

\begin{equation}
F_{\text{QED}}(T)=-\left[ \frac{\pi ^{3}(kT)^{4}\gamma \left( 3\omega
_{0}^{2}-\gamma ^{2}\right) }{45\hbar ^{3}\omega _{0}^{6}}+\frac{8\pi
^{5}(kT)^{6}\gamma \left( 5\omega _{0}^{4}-5\gamma ^{2}\omega
_{0}^{2}+\gamma ^{4}\right) }{315\hbar ^{5}\omega _{0}^{10}}+\cdots
\right], 
\end{equation}

\begin{equation}
S_{\text{QED}}(T)=k\left[ \frac{4\pi ^{3}(kT)^{3}\gamma \left( 3\omega
_{0}^{2}-\gamma ^{2}\right) }{45\hbar ^{3}\omega _{0}^{6}}+\frac{16\pi
^{5}(kT)^{5}\gamma \left( 5\omega _{0}^{4}-5\gamma ^{2}\omega
_{0}^{2}+\gamma ^{4}\right) }{105\hbar ^{5}\omega _{0}^{10}}\cdots
\right], 
\end{equation}

\begin{equation} U_{\text{QED}}(T)=\frac{\pi ^{3}(kT)^{4}\gamma \left(
3\omega _{0}^{2}-\gamma ^{2}\right) }{15\hbar ^{3}\omega
_{0}^{6}}+\frac{8\pi ^{5}(kT)^{6}\gamma \left( 5\omega _{0}^{4}-5\gamma
^{2}\omega _{0}^{2}+\gamma ^{4}\right) }{63\hbar ^{5}\omega
_{0}^{10}}+\cdots, 
\end{equation} and

\begin{equation} C_{\text{QED}}(T)=k\left[ \frac{4\pi ^{3}(kT)^{3}\gamma
\left( 3\omega _{0}^{2}-\gamma ^{2}\right) }{15\hbar ^{3}\omega
_{0}^{6}}+\frac{16\pi ^{5}(kT)^{5}\gamma \left( 5\omega _{0}^{4}-5\gamma
^{2}\omega _{0}^{2}+\gamma ^{4}\right) }{21\hbar ^{5}\omega
_{0}^{10}}+\cdots
\right]. 
\end{equation}  We note that, in the large cut-off limit \cite{ford7},
$\gamma=\omega^{2}_{0}\tau_{e}$, where $\tau_{e}$ is given in (2.15).
In this limit and with $\gamma<<\omega_{0}$, we have a check in that the
leading terms in the free energy and the entropy agree with the results
obtained earlier by us \cite{ford4}.

\subsection{High Temperature Expansion ($kT>>\hbar\omega_{0}$)}

With the high temperature expansion (3.9) for $F_{Ohmic}$ we find from the
general expression (4.4)

\begin{equation}
F_{QED}(T)=-kT\log\frac{kT}{\hbar\omega_{0}}+\frac{\pi(kT)^{2}\gamma}
{6\hbar\omega^{2}_{0}}+\cdots,
\end{equation}

\begin{equation}
S_{QED}(T)=k\left\{(\log\frac{kT}{\hbar\omega_{0}}+1)-\frac{\pi(kT)\gamma}
{3\hbar\omega^{2}_{0}}+\cdots\right\},
\end{equation}

\begin{equation}
U_{QED}(T)=kT-\frac{\pi\gamma}
{3\hbar\omega^{2}_{0}}(kT)^{2}+\cdots,
\end{equation} and

\begin{equation}
C_{QED}(T)=k\left\{\frac{4\pi
^{3}(kT)^{3}\gamma \left( 3\omega _{0}^{2}-\gamma ^{2}\right) }{15\hbar
^{3}\omega _{0}^{6}}+\cdots\right\}.
\end{equation}  We note that these results agree with the corresponding
results in \cite{ford1}.

\section{Zero-point energy}

\label{sec:five}

Since $F=U+TS$, the zero-point free energy is always identical with the
zero-point energy. The zero-point free energy is obtained by replacing $
f(\omega ,T)\rightarrow \hbar \omega /2$ in the formula (\ref{eq9}). The
resulting expression diverges for the QED model, whatever the cutoff. For
the single relaxation time model it is finite for finite relaxation time,

\begin{eqnarray}
\left( F\right) _{zero-point} &=&\frac{\hbar }{2\pi }\int_{0}^{\infty
}d\omega \{-\frac{\Omega \omega }{\omega ^{2}+\Omega ^{2}}+\frac{\Omega
^{\prime }\omega }{\omega ^{2}+\Omega ^{\prime 2}}+\frac{z_{1}\omega }{
\omega ^{2}+z_{1}^{2}}+\frac{z_{1}^{\ast }\omega }{\omega ^{2}+z_{1}^{\ast 2}
}\}  \notag \\
&=&\frac{\hbar }{2\pi }\{\Omega \log \Omega -\Omega ^{\prime }\log \Omega
^{\prime }-z_{1}\log z_{1}-z_{1}^{\ast }\log z_{1}^{\ast }\}  \notag \\
&=&\frac{\hbar }{2\pi }\{\Omega ^{\prime }\log \frac{\Omega ^{\prime
}+\gamma }{\Omega ^{\prime }}+\gamma \log \frac{\Omega ^{\prime }+\gamma }{
\omega _{0}}+2\omega _{1}\arccos \frac{\gamma }{2\omega _{0}}\}.
\end{eqnarray}

In the Ohmic limit this is logarithmically divergent,

\begin{equation}
\left( F\right) _{\text{zero-point}}\sim \frac{\hbar }{2\pi }\{\gamma
(1-\log \omega _{0}\tau )+2\omega _{1}\arccos \frac{\gamma }{2\omega _{0}}\}.
\end{equation}

\section{Conclusions}

Motivated by the fact that environmental effects play an important role in
many topical areas of physics, where dissipation and fluctuation effects
often play a significant role, we have presented an \underline{exact}
calculation of quantum thermodynamic functions for an oscillator in an
arbitrary heat bath at arbitrary temperatures.  Explicit results were
obtained for both high and low temperatures.  Since we are dealing
with \underline{non-additivity of entropies} \cite{ford4}, we use a method
based on (2.8), which is an exact result for the free energy of an
oscillator which takes into account interaction effects.  In the
Introduction, we have already given examples of its application
\cite{li,ford4,ford5}.  However, there are many other possible topics
where such results are likely to be applicable.  For example, Jordan and
Buttiker \cite{jordan} have demonstrated the relation between entanglement
(due to the heat bath) and energy fluctuations and concluded that large
entanglement implies large energy fluctuations.  Since their work was
confined to zero temperature, it would be of interest to extend it to
non-zero temperatures.  In a similar vein, the decrease of the coherence
length of an Aharonov-Bohm-like interferometer due to interaction with the
environment was examined but again it was confined to zero temperature
\cite{ratch}. 

Finally, we turn to a very different area where thermodynamic
considerations play a vital role i.e. the study of black holes. 
Following the remarkable results of Bekenstein and Hawking \cite{bek},
there has been continuing interest in developing a microscopic theory for
the entropy of a black hole and, in particular, the fact that it depends
on the area of the event horizon.  As an example, we mention the work of
Bombelli et al. \cite{bomb} and Srednicki \cite{sred}, where the use of
partial traces and reduced density matrices played a crucial role.  Since,
in general, such techniques lead to results different from those obtained
by the method discussed above, we feel that it would be worthwhile to
apply our approach to the study of the thermodynamic properties of black
holes.

\newpage

\appendix

\section{The Stieltjes $\mathbf{J}$\textbf{-function}}

The Stieltjes $J$-function is introduced by the integral:\cite{wall}

\begin{equation}
J(z)=-{\frac{1}{\pi }}\int_{0}^{\infty }dt\log (1-e^{-2\pi t}){\frac{z}{
z^{2}+t^{2}}},\qquad \mathrm{Re}\{z\}>0.  \label{77.1}
\end{equation}  The imaginary axis is a \textquotedblleft natural
boundary\textquotedblright\ of $J(z)$. That is, the analytic continuation of 
$J(z)$ into the left half plane is not given by the analytic continuation of
the integral.

This analytic continuation is based on the identity:\cite{wall}

\begin{equation}
J(z)=\log [\Gamma (z+1)]-\log \sqrt{2\pi }-(z+{\frac{1}{2}})\log (z)+z.
\label{77.2}
\end{equation} Since $\Gamma (z)$ is analytic in the entire plane except for poles at
$z=0,-1,-2,\cdots$, we can use this form throughout the $z$-plane cut along
the negative real axis. It is then a simple matter to show that the
continued $J$-function is given by 

\begin{equation}
J(ze^{\pm i\pi })=-J(z)-\log (1-e^{\mp 2\pi iz}),\qquad \mathrm{Re}(z)>0.
\label{77.3}
\end{equation}  For $\left\vert z\right\vert <1$, we have the expansion
\cite{bateman_htf1}

\begin{equation}
J(z)=-\log\sqrt{2\pi}-(z+1/2)\log~z+z-\gamma_{E}~z+
\sum_{n=2}^{\infty}{\frac{(-)^{n}\zeta(n)}{n}}~z^{n}, \label{77.4}
\end{equation}

\noindent where $\gamma_{E}$=0.5772157 is Euler's constant and $\zeta
(n)$ is the Riemann $\zeta$-function. For large $z$ we have the asymptotic
expansion:\cite{bateman_htf1} 

\begin{equation}
J(z)=\sum_{n=0}^{\infty }\frac{B_{2n+2}}{(2n+1)(2n+2)}\frac{1}{z^{2n+1}},
\label{A.5}
\end{equation} where the Bernoulli numbers are 

\begin{eqnarray}
B_{2} &=&{\frac{1}{6}},\quad B_{4}=-\frac{1}{30},\quad B_{6}={\frac{1}{42}}
,\quad B_{8}=-{\frac{1}{30}},  \nonumber \\
B_{10} &=&{\frac{5}{66}},\quad B_{12}=-{\frac{691}{2730}},\quad B_{14}={
\frac{7}{6}},\quad B_{16}=-{\frac{3617}{510}},  \nonumber \\
B_{18} &=&{\frac{43867}{798}},\quad B_{20}=-{\frac{174611}{330}},\quad
B_{22}={\frac{854513}{138}},\cdots .  \label{77.7}
\end{eqnarray}   Very useful for numerical computation is the Lanczos
formula \cite{lanczos,press}: 

\begin{equation}
J(z)=(z+\frac{1}{2})\log {\frac{z+\gamma +\frac{1}{2}}{z}}-\gamma -\frac{1}{2}+\log
\{d_{0}+\sum_{n=1}^{N}{\frac{d_{n}}{z+n}}\},\qquad Re~z>0,
\label{77.15}
\end{equation} where, for $N=6,\ \gamma =5$, and 

\begin{eqnarray}
d_{0} &=&1.000000000190015,\quad d_{1}=76.18009172947146,  \nonumber \\
d_{2} &=&-86.50532032941677,\quad d_{3}=24.01409824083091,  \nonumber \\
d_{4} &=&-1.231739572450155,\quad d_{5}=0.001208650973866179,  \nonumber \\
d_{6} &=&-0.000005395239384953.  \label{77.16}
\end{eqnarray}  The numerical error is small (less than a part per
billion) everwhere in the right half plane.

\end{document}